\documentclass[12pt,preprint]{aastex}

\newcommand{\bdv}[1]{\mbox{\boldmath$#1$}}

\def\au{{\rm AU}} 
 
\def\kms{{\rm km}\,{\rm s}^{-1}}

\def\rel{{\rm rel}}

\def\e{{\rm E}}
\def\bpi{{\bdv\pi}}
\def\bmu{{\bdv\mu}}
\def\btheta{{\bdv\theta}}

\begin{document}
\title{OGLE-2011-BLG-0417: A Radial Velocity Testbed for Microlensing}

\author{
Andrew~Gould\altaffilmark{1},
In-Gu~Shin\altaffilmark{2},
Cheongho~Han\altaffilmark{2},
Andrzej~Udalski\altaffilmark{3},
Jennifer~C.~Yee\altaffilmark{1}
 }
\altaffiltext{1}{Department of Astronomy, Ohio State University,
140 W.\ 18th Ave., Columbus, OH 43210, USA; 
gould@astronomy.ohio-state.edu}
\altaffiltext{2}{Department of Physics, Chungbuk National University, 
Cheongju 361-763, Korea}
\altaffiltext{3}{Warsaw University Observatory, Al. Ujazdowskie 4, 
00-478 Warszawa, Poland}

\begin{abstract}

Microlensing experiments are returning increasingly detailed information
about the planetary and binary systems that are being detected,
far beyond what was originally expected.  In several cases the lens mass
and distance are measured, and a few very special cases have yielded
complete 8-parameter Kepler solutions, i.e., the masses of both components, 
five Kepler invariants and the phase.  We identify one such case that
is suitable for a precision test that could be carried out by comparing
Doppler (RV) measurements with the predictions from the microlensing solution.
The lens primary is reasonably bright ($I=16.3$, $V=18.0$) and is expected
to have a relatively large RV semi-amplitude ($K\sim 6.35\,\kms$).

\end{abstract}


\section{{Introduction}
\label{sec:intro}}

When microlensing planet searches were first proposed
\citep{liebes64,mao91,gouldloeb92} it was not expected that much information
would be extracted about the individual planets that were detected.
\citet{gouldloeb92} pointed out that the planet-host mass ratio $q$ and the 
planet-host projected separation $s$ in units of the Einstein radius can
be measured, but no other parameter measurements were mentioned.  
By contrast, Doppler
(RV) detections routinely return six independent parameters out
of the eight possible, i.e., the masses of the two bodies, plus six phase-space
coordinates.  The latter are usually parameterized by five Kepler-orbit
invariants plus the orbital phase.  After one includes a spectroscopic
determination of the host mass, RV lacks only the orientation of
the orbit on the plane of the sky (which is generally of little interest)
and resolution of the famous $m\sin i$ degeneracy, where $m$ is the
planet mass and $i$ is the orbital inclination.

However, over the course of two decades, our understanding of what
can be extracted from microlensing planet and binary detections has gradually
expanded to the point that now, incredibly, sometimes all eight parameters
are reported.  There only remains a single discrete degeneracy:
the sign of the radial velocity.

First, if the source
passes over a caustic caused by the lens, one can measure the angular
size of the source relative to the Einstein ring, $\rho=\theta_*/\theta_\e$
\citep{gould94,witt94,nemiroff94}.  The very first microlensing planet
OGLE-2003-BLG-235Lb showed such ``finite-source effects'' \citep{ob03235},
and indeed the majority of subsequent microlensing planets have as well.
Since $\theta_*$ can be measured from an instrumental color magnitude
diagram (CMD) of the event \citep{yoo04} and such color data are routinely
taken, the majority of events also yield $\theta_\e$.  \citet{gould92}
pointed out that the ``microlens parallax'' $\pi_\e$ could in principle
be measured from lightcurve distortions due to Earth's motion, and that
if both $\pi_\e$ and $\theta_\e$ could be measured, then so could the
lens mass.  Here,
\begin{equation}
\theta_\e^2 = \kappa M\pi_\rel;
\quad
\pi_\e^2 = {\theta_\e\over \kappa M}; 
\quad \kappa\equiv {4 G\over c^2\,\au}\simeq 8.1\, {{\rm mas}\over M_\odot},
\label{eqn:thetaepie}
\end{equation}
$M$ is the total mass of the lens, and $\pi_\rel\equiv \pi_L-\pi_S$ 
is the lens-source relative parallax.  Already, 
the second microlensing planet OGLE-2005-BLG-071 \citep{ob05071}
showed such parallax distortions, and thus became the first microlensing 
planet with a 
mass measurement \citep{dong09}.  A substantial minority of subsequent
microlensing planets yielded microlens parallax measurements as well.
Since, the source distance (and so 
$\pi_S = \au/D_S$) is usually known quite well, Equation (\ref{eqn:thetaepie})
also yields the lens distance $D_L$ and so also the physical projected
separation $r_\perp = s\theta_\e D_L$.  Since the microlens parallax
measurement is actually described by a vector $\bpi_\e$ \citep{gould04} whose
direction is that of the lens-source relative motion, the projected
separation ${\bf r}_\perp$ in fact contains two phase-space coordinates.

This still leaves four phase-space coordinates to be determined for
a complete orbital solution.  It was an enormous surprise when the first orbital
motion was derived from the microlensing event MACHO-BLG-97-041
\citep{mb97041} because
the perturbations due to the lens structure typically last a few days
while the orbital periods are expected to be several years.  Even then
the measurement was considered highly unique, due to accidental geometry.
Moreover, even in that exceptionally favorable circumstance, only
two additional parameters were measured: the time rate of change of
the binary separation $ds/dt$, which affects the shape of the caustic;
and the time rate of change of the angular orientation of the lens axis,
$\omega=d\alpha/dt$, which affects the orientation of the caustic.
Hence, in this case, the four phase-space coordinates in the plane
of the sky were measured, and the two radial coordinates were not.
This seems natural because microlensing is sensitive to the mass distribution
projected along the line of sight.  Several planetary events have yielded
such measurements, or partial measurements, the first two being
OGLE-2005-BLG-071 \citep{dong09} and OGLE-2006-BLG-109 \citep{ob06109,ob06109b}.

Even as these measurements began to accumulate it was regarded as
virtually impossible to obtain information about the orbit in the
third direction, which would be the only way to derive the invariants of
a Kepler orbit.  And indeed, no such measurements have yet
been made for planetary events.  However, just as it is possible to
use Kepler's Second Law to deproject astrometric data on visual binaries
to obtain a full orbit, it should in principle be possible to deproject
microlensing-binary orbits.  The challenge is, again, that the perturbations
are generally short compared to an orbital period.
Nevertheless, \citet{ob09020} analyzed the binary-lens event
OGLE-2009-BLG-020 and found that two additional parameters were required
to describe the event, $s_z$ and $\gamma_z$, the position and time rate
of change of the binary separation in the radial direction.  That is,
all six phase-space coordinates were needed.  Now, in that particular
case, the fits to these parameters were highly correlated (see their
Figure 3).  However, \citet{ob09020} took this opportunity to work
out the relations of microlensing parameters and Kepler parameters
(see their Appendices A and B).   Then, making use of this formalism,
\citet{ob05018,ob110417} were able to obtain complete (8-parameter)
Kepler solutions for three events, OGLE-2005-BLG-018, MOA-2011-BLG-090,
and OGLE-2011-BLG-0417.  In this paper, we will show how the complete solution
for OGLE-2011-BLG-0417 allows a direct test of the microlensing model.

\section{{Past Tests of Microlensing}
\label{sec:tests}}

The importance of these measurements goes far beyond what they tell
us about the individual systems.  Microlensing events are famously
``unrepeatable''.  This does not mean that the results cannot be
corroborated, but it does pose challenges.  

For example, few would doubt the basic interpretation of either
of the two-planet systems discovered by microlensing,
OGLE-2006-BLG-109Lb,c \citep{ob06109,ob06109b} and OGLE-2012-BLG-0026Lb,c
\citep{ob120026}.  All the features of these observed
events can be understood in terms of well-established (if specialized)
microlensing principles.  Several of the major features are corroborated
by overlapping data sets.  Nevertheless, there are aspects of the
modeling of both events that yield nominally high signal-to-noise ratio
parameter measurements without corresponding ``visible and unambiguous''
lightcurve features.  In particular, both events yield microlens
parallax measurements that have obvious signatures for only one of the
two components of the microlens parallax vector 
$\bpi_\e=(\pi_{\e,\parallel},\pi_{\e,\perp})$.  That is, the component of
$\bpi_\e$ parallel to the projected position of the Sun ($\pi_{\e,\parallel}$)
leads to an obvious asymmetry in the light curve \citep{gmb94}, while
the signature of $\pi_{\e,\perp}$ is entangled with many other lightcurve
parameters, in particular $\omega$, 
the component of lens orbital motion perpendicular to the binary axis
\citep{mb09387,ob09020}.  Hence, it is not at all obvious that 
$\bpi_\e$, and so $\pi_\e=|\bpi_\e|$, and thus $M$ and $D_L$ are being
measured correctly.

In most cases, tests of these subtle higher-order microlensing parameter
measurements are impossible.  Fortunately, however, there are a few
rare cases for which they are possible.  Moreover, there are no obvious
reasons that the events that can be tested are more likely to have
correctly measured microlensing parameters than those that cannot.

The number of such tests that have been carried out in the past is
actually quite small.  In fact, there are just two.  However, both
were spectacular successes.   The first was MACHO-LMC-5, an event
that occurred in 1993, whose lightcurve yielded a measurement of $\bpi_\e$
but not $\theta_\e$.  However, $\Delta t= 6.3\,$yr later, \citet{alcock01}
imaged the lens and source using the {\it Hubble Space Telescope (HST)}
and thereby found the lens-source relative proper motion
$\bmu=\Delta\btheta/\Delta t$, where $\Delta\btheta$ is the vector separation
between the lens and source stars.
When combined with the Einstein timescale $t_\e$ measured from the event, this
yielded $\theta_\e = \mu t_\e$.  These measurements allowed for 
two tests.  First, the
photometry of the lens should have been consistent with the
mass and distance inferred from the measurements of $\pi_\e$ (lightcurve) and
$\theta_\e$ (lightcurve plus {\it HST} astrometry).  
Second, the direction of $\bmu$ ({\it HST} astrometry) should have been the
same as that of $\bpi_\e$ (lightcurve).  
In fact, the observations appeared to fail both
tests.  Subsequently, however, \citet{gould04} found a discrete degeneracy
in the parallax solution, with the other solution yielding consistency
for both tests.  Moreover, \citet{drake04} obtained a trigonometric
parallax for the lens and confirmed the alternative
microlensing parallax derived by \citet{gould04}.

The second tested event was OGLE-2006-BLG-109, the first two-planet event.
This event yielded both $\bpi_\e$ and the transverse orbital parameters
$ds/dt$ and $\omega$.  Recall that $\pi_{\e,\perp}$ can be entangled with
$\omega$ as well as other parameters.  The lens flux predicted on the
basis of the mass and distance derived from $\pi_\e$ and $\theta_\e$
was far smaller than the blended light superposed on the source.  This
could have either been because of errors in $\pi_\e$ and/or $\theta_\e$,
or because there was additional light in the aperture, i.e., not related
to either the source or lens of the event.  \citet{ob06109b} obtained
AO images using the Keck telescope and found that the blended light
was clearly displaced from the event and that the ``object'' at the
location of the event was consistent with the combined light from the
source and lens, as predicted by the microlens solution.

\section{{A New Test}
\label{sec:newtest}}

However, to date, there has never been a test of microlensing
orbital-parameter measurements.  We here propose such a test.
Of the three binary events with complete solutions, OGLE-2011-BLG-0417
is by far the best candidate.  OGLE-2005-BLG-018 has an extremely
bright source star, which would preclude making measurements
of the lens star until the two separate by at least several hundred mas,
many decades from now.  The brighter component of MOA-2011-BLG-090L is
expected to be $I\ga 20.5$, making it virtually impossible to monitor
at the required RV precision.

However, OGLE-2011-BLG-0417L is so bright, it is easily identified as the
``blended light'' in the event CMD (see Figure 4 of \citealt{ob110417}).
Indeed, the fact that this blended-light point sits right on the 
``disk main sequence'' or ``reddening track'' at roughly the position
expected for the primary component of the lens, is already an indication
that the microlensing solution is basically correct.  But because
the lens is bright, $I_L\sim 16.3$, $V_L\sim 18.0$, it is also feasible
to test whether the orbit as determined from the microlensing solution
is correct.

Table~\ref{tab:ulensparm} shows the eight Kepler parameters determined
from microlensing (total mass, mass ratio, 5 Kepler invariants, and time
of periastron), plus the lens distance, which is also determined.
These are derived from the underlying Markov chains used by \citet{ob110417}
except that we have adopted the new bulge clump giant calibration of
\citet{nataf12} and
also have allowed for the 5\% error in $\theta_*$, which was not previously
included.  Note that in order to express the results as precisely as possible,
the correlation coefficients are displayed, as well as the errors.

Table~\ref{tab:rvparm} shows our predictions (and covariance matrix)
for the five Kepler parameters
that can be measured by RV (velocity semi-amplitude, period, eccentricity,
argument of periastron, time of periastron) as well as the primary mass
and system distance, which can both be estimated from the spectrum
(in the latter case augmented by calibrated photometry).  In this case,
it is especially important to state the correlation coefficient between
the period and time of periastron, since the particular orbit when
$t_p$ will be measured is not known in advance.

\begin{figure}
\plotone{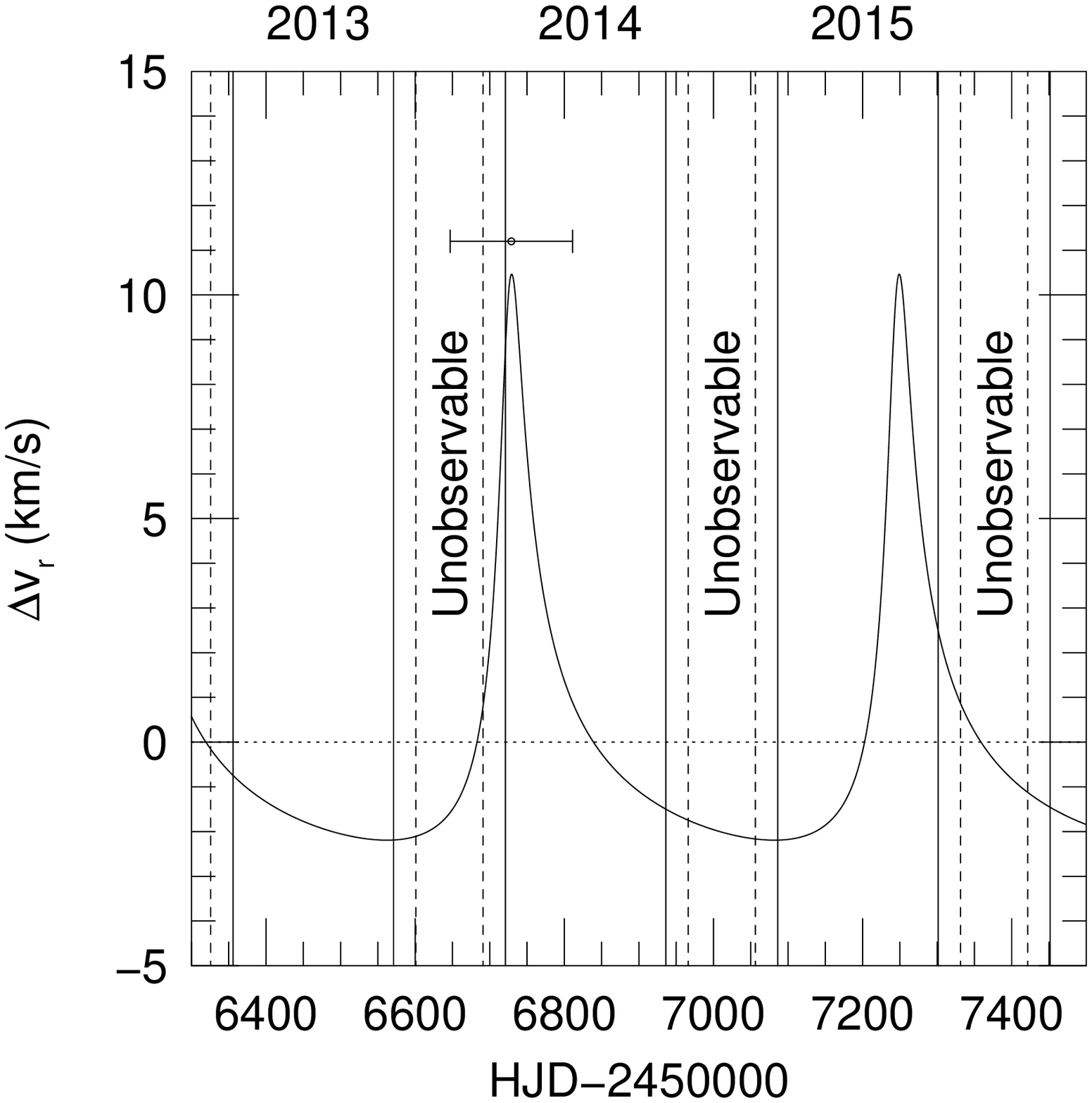}
\caption{\label{fig:rv}
Predicted RV curve for the primary of OGLE-2011-BLG-0417L, based
on the analysis of \citet{ob110417}.  Dashed lines indicate $\pm 45\,$days
from the winter Solstice, when observations are virtually impossible
from Earth.  Solid lines show a more conservative exclusion
window of $\pm 75\,$days.  Note that microlensing predictions are
intrinsically ambiguous as to the sign of the RV curve.
While the {\it form} of this curve is well-predicted, the {\it phase}
is gradually being lost: the error bar at the first periastron indicates 
the phase error at that time.  RV observations can easily recover the
phase.
}
\end{figure}

Figure~\ref{fig:rv} shows the predicted RV curve over the next
several years.  It is virtually impossible to make observations
within 45 days to the Winter Solstice.  The resulting exclusion
zones are marked in dashed lines.  We also mark a more conservative
window of $\pm 75\,$days from the Winter Solstice.  Note that
while the {\it form} of the RV curve is well-predicted,
the {\it phase} is gradually being lost.  The error bar in the
figure indicates the uncertainty in the time of the first
periastron.  Because of the high eccentricity, this phase
can be recaptured by a few judiciously placed RV observations.

The predicted RV amplitude $(K=6.35\pm 0.34,\kms)$ is relatively high
and should be measurable with good precision, despite the relatively
faint (for typical RV work) target, $I\sim 16.3$, $V\sim 18.0$.
One complicating factor is that (as with all microlensing events),
the lens is superposed on the source, which in this case
is a clump giant.
However, because the source is almost 10 times more distant and
is seen through $\Delta E(V-I)\sim 1$ mag more reddening, the lens is
actually brighter than the source by about 0.4 mag in $I$ and and 1.3 mag
in $V$.  See Figure~4 of \citet{ob110417}.
Moreover, given that the lens is in the Galactic Disk whereas
the source is in the Bulge, they are likely to have RVs that differ
by several tens of $\kms$.

\section{{Conclusion}
\label{sec:conclude}}

We have proposed a rigorous test of microlensing parameters via
RV measurements of OGLE-2011-BLG-0417L, a binary lens with a complete
orbital solution.  The expected RV semi-amplitude is $6.35\pm 0.34\,\kms$,
which should be precisely measurable despite the fact that the
lens primary is relatively faint ($I=16.3$, $V=18.0$) and is superposed
on a somewhat fainter star that was the source star in the microlensing
event.
We have presented both the predicted values and error bars of all
seven quantities that are measurable with RV, including 5 Kepler parameters
(or parameter combinations), the primary mass (from spectroscopic typing) 
and the distance (from combined spectroscopy and photometry).

This would be only the third precision test of microlensing by 
external measurements, and would be by far the most exacting, since
many more parameter determinations would be tested in a much more
complicated system.



\acknowledgments

AG was supported by NSF grant AST 1103471.
Work by CH was supported by Creative Research Initiative
Program (2009-0081561) of National Research Foundation
of Korea.
The OGLE project has received funding from the European Research Council
under the European Community's Seventh Framework Programme
(FP7/2007-2013) / ERC grant agreement no. 246678 to AU.
Work by JCY was supported by an NSF Graduate Research Fellowship under
Grant No. 2009068160.

\begin{table}
\caption{\label{tab:ulensparm} \sc Microlens Measurements}
\vskip 1em
\begin{tabular}{@{\extracolsep{0pt}}lrrrrrrrrr}
\hline
\hline
   & $M_{\rm tot}$ & $M_2/M_1$ & $P$ & $e$ & $i$ & $\omega$ & $\Omega$ & $t_{\rm peri}$ & $D_L$  \\ \hline
 & ($M_\odot$) & & (yr) & & (deg) & (deg) & (deg) & (HJD) & (kpc)\\ \hline \hline
Value & $  0.677$ & $  0.292$ & $  1.423$ & $  0.688$ & $ 60.963$ & $341.824$ & $125.374$ & $5686.344$ & $  0.951$\\
Error & $  0.047$ & $  0.003$ & $  0.113$ & $  0.027$ & $  1.554$ & $  2.655$ & $  1.649$ & $  6.960$ & $  0.058$\\
 \hline
C.C.  & $  1.000$ & $ -0.204$ & $  0.101$ & $  0.511$ & $ -0.024$ & $ -0.008$ & $  0.511$ & $  0.133$ & $ -0.065$\\
C.C.  & $ -0.204$ & $  1.000$ & $ -0.055$ & $ -0.150$ & $  0.161$ & $ -0.065$ & $ -0.302$ & $ -0.015$ & $ -0.136$\\
C.C.  & $  0.101$ & $ -0.055$ & $  1.000$ & $ -0.118$ & $  0.523$ & $  0.484$ & $  0.595$ & $ -0.756$ & $  0.791$\\
C.C.  & $  0.511$ & $ -0.150$ & $ -0.118$ & $  1.000$ & $ -0.217$ & $ -0.247$ & $  0.151$ & $  0.485$ & $  0.211$\\
C.C.  & $ -0.024$ & $  0.161$ & $  0.523$ & $ -0.217$ & $  1.000$ & $ -0.257$ & $  0.141$ & $ -0.781$ & $  0.093$\\
C.C.  & $ -0.008$ & $ -0.065$ & $  0.484$ & $ -0.247$ & $ -0.257$ & $  1.000$ & $  0.667$ & $ -0.013$ & $  0.518$\\
C.C.  & $  0.511$ & $ -0.302$ & $  0.595$ & $  0.151$ & $  0.141$ & $  0.667$ & $  1.000$ & $ -0.141$ & $  0.469$\\
C.C.  & $  0.133$ & $ -0.015$ & $ -0.756$ & $  0.485$ & $ -0.781$ & $ -0.013$ & $ -0.141$ & $  1.000$ & $ -0.367$\\
C.C.  & $ -0.065$ & $ -0.136$ & $  0.791$ & $  0.211$ & $  0.093$ & $  0.518$ & $  0.469$ & $ -0.367$ & $  1.000$\\
\hline
\end{tabular}
\end{table}

\begin{table}
\caption{\label{tab:rvparm} \sc Predictions for RV Measurements}
\vskip 1em
\begin{tabular}{@{\extracolsep{0pt}}lrrrrrrr}
\hline
\hline
   & $K$ & $P$ & $e$ & $\omega$ & $t_{\rm peri}$ & $M_1$ & $D_L$  \\ \hline
& $({\rm km\,s^{-1}})$ & (yr) & & (deg) & (HJD) & ($M_\odot$) & (kpc)\\ \hline \hline
Value & $  6.352$ & $  1.423$ & $  0.688$ & $341.824$ & $  5686.344$ & $  0.524$ & $  0.951$\\
Error & $  0.340$ & $  0.113$ & $  0.027$ & $  2.655$ & $  6.960$ & $  0.036$ & $  0.058$\\
 \hline
C.C.  & $  1.000$ & $ -0.365$ & $  0.838$ & $ -0.473$ & $  0.503$ & $  0.693$ & $ -0.267$\\
C.C.  & $ -0.365$ & $  1.000$ & $ -0.118$ & $  0.484$ & $ -0.756$ & $  0.101$ & $  0.791$\\
C.C.  & $  0.838$ & $ -0.118$ & $  1.000$ & $ -0.247$ & $  0.485$ & $  0.511$ & $  0.211$\\
C.C.  & $ -0.473$ & $  0.484$ & $ -0.247$ & $  1.000$ & $ -0.013$ & $ -0.008$ & $  0.518$\\
C.C.  & $  0.503$ & $ -0.756$ & $  0.485$ & $ -0.013$ & $  1.000$ & $  0.133$ & $ -0.367$\\
C.C.  & $  0.693$ & $  0.101$ & $  0.511$ & $ -0.008$ & $  0.133$ & $  1.000$ & $ -0.065$\\
C.C.  & $ -0.267$ & $  0.791$ & $  0.211$ & $  0.518$ & $ -0.367$ & $ -0.065$ & $  1.000$\\
\hline
\end{tabular}
\end{table}


\begin{thebibliography}{99}

\bibitem[Albrow et al.(2000)]{mb97041}
Albrow, M.D., Beaulieu, J.-P.. Caldwell, J.A.R. 2000, \apj, 534, 894

\bibitem[Alcock et al.(2001)]{alcock01}
Alcock, C., Allsman, R.A., Alves, D.R. et al. 2001, Nature, 414, 617

\bibitem[Batista et al.(2011)]{mb09387} 
Batista, V., Gould, A., Dieters, S. et al.\ 2011, \aap, 529,110

\bibitem[Bennett et al.(2010)]{ob06109b}
Bennett, D.P., Rhie, S.H., Nikolaev, S. et al. 2010, \apj, 713, 837


\bibitem[Bond et al.(2004)]{ob03235} 
Bond, I.A., Udalski, A., Jaroszy\'nski, M.\  et al. 2004, \apj, 606, L155

\bibitem[Drake et al.(2004)]{drake04}
Drake, A.J., Cook, K.H., \& Keller, S.C. 2004, \apj 607 L29

\bibitem[Han et al.(2013)]{ob120026} 
Han, C., Udalski, A., Choi, J.-Y.  et al. 2013, \apj, 762, L28

\bibitem[Dong et al.(2009)]{dong09} 
Dong, S., Gould, A., Udalski, A. et al. 2009, \apj, 695, 970

\bibitem[Gaudi et al.(2008)]{ob06109}
Gaudi, B.S., Bennett, D.P., Udalski, A. et al.\ 2008, Science, 319, 927

\bibitem[Gould(1992)]{gould92} 
Gould, A. 1992, \apj, 392, 442

\bibitem[Gould(1994)]{gould94} 
Gould, A. 1994, \apj, 421, L71

\bibitem[Gould(2004)]{gould04} Gould, A.  2004, \apj, 606, 319

\bibitem[Gould \& Loeb(1992)]{gouldloeb92} 
Gould, A., \& Loeb, A.\ 1992, \apj, 396, 104

\bibitem[Gould et al.(1994)]{gmb94} 
Gould, A., Miralda-Escud\'e \&  Bahcall, J.N. 1994, \apj, 423, L105

\bibitem[Liebes(1964)]{liebes64} Liebes, S.\ 1964, Physical Review, 133, 835

\bibitem[Mao \& Paczy\'nski(1991)]{mao91}
Mao, S. \& Paczy\'nski, B.\ 1991, \apj, 374, L37

\bibitem[Nataf et al.(2012)]{nataf12} Nataf, D.M., Gould, A., 
Fouqu\'e, P. et al. 2012, \apj, submitted, arXiv:1208:1263

\bibitem[Nemiroff \& Wickramasinghe(1994)]{nemiroff94}
Nemiroff, R.J., \& Wickramasinghe, W.A.D.T.\ 1994, \apj, 424, L21

\bibitem[Shin et al.(2011)]{ob05018}
Shin, I.-G., Udalski, A., Han, C. et al.\ 2011, \apj, 735, 85

\bibitem[Shin et al.(2012)]{ob110417}
Shin, I.-G., Han, C., Choi, J.-Y., et al.\ 2012, \apj, 755, 91

\bibitem[Skowron et al.(2011)]{ob09020}
Skowron, J., Udalski, A., Gould, A et al.\ 2011, \apj, 738, 87

\bibitem[Udalski et al.(2005)]{ob05071}
Udalski, A., Jaroszy\'nski, M., Paczy\'nski, B., et al.\ 2005, \apj, 628, L109

\bibitem[Witt \& Mao(1994)]{witt94}
Witt, H.J., \& Mao, S.\ 1994, \apj, 429, 66

\bibitem[Yoo et al.(2004)]{yoo04} 
Yoo, J., DePoy, D.L., Gal-Yam, A. et al.\ 2004, \apj, 603, 139

\end{thebibliography}
\end{document}